\documentclass[manuscript]{aastex}
\usepackage[usenames]{color}
\usepackage{amsmath}
\usepackage{graphicx}
\usepackage{amssymb}

\shorttitle{The  effect of influential points on the Levitt law}
\shortauthors{Garc\'ia-Varela et al.}

\begin{document}


\title{The influential effect of blending, bump,  
changing period and eclipsing Cepheids on the Leavitt law}



\author{A. Garc\'ia-Varela, J. R. Mu\~noz, B. E. Sabogal}
\affil{Universidad de los Andes, Departamento de F\'{\i}sica,
  Cra. 1 No. 18A-10, Bloque Ip, A.A. 4976, Bogot\'a, Colombia}     

  \and
\author{S. Vargas Dom\'inguez}    
\affil{Universidad Nacional de Colombia - Sede Bogot\'a - 
Facultad de Ciencias - Observatorio Astron\'omico - Carrera 45 No. 26-85, 
Bogot\'a - Colombia}  
 
  \and

\author{J. Mart\'inez}
\affil{Universidad de los Andes, Departamento de Ingenier\'ia Industrial, Edificio ML, Cra 1 Este No 19A - 40, Bogot\'a, Colombia}
      
\email{josegarc@uniandes.edu.co, jr.munoz2198@uniandes.edu.co, 
bsabogal@uniandes.edu.co, svargasd@unal.edu.co, 
j.martinez144@uniandes.edu.co}




\begin{abstract}
The investigation of the non-linearity of the Leavitt law is a
topic that began more than seven decades ago, when some 
of the studies in this field found that the Leavitt 
law has a break at about ten days.\\
The goal of this work is to investigate a possible statistical 
cause of this non-linearity. By applying linear regressions to OGLE-II and OGLE-IV data, we find that,
in order to obtain the Leavitt law by using linear regression,
robust techniques to deal with influential points and/or outliers 
are needed instead of the ordinary least-squares regression traditionally used. 
In particular, by using $M$-  and $MM$-regressions we establish firmly and 
without doubts the linearity of the Leavitt law in the Large Magellanic Cloud, without rejecting 
or excluding Cepheid data from the analysis. This implies 
that light curves of Cepheids suggesting blending,  bumps, 
eclipses or  period changes, do not affect the Leavitt law for this galaxy. For
the SMC, including this kind of Cepheids, it is not possible to find 
an adequate model, probably due to the geometry 
of the galaxy. In that case, a possible influence of these stars could exist. 
\end{abstract}





\section{Introduction}
By studying variable stars in the Small Magellanic Cloud (SMC), 
Henrietta Leavitt discovered a linear relation between the 
pulsation period and magnitudes of $25$ Cepheids, in the sense of that the 
brightest Cepheids have longer periods \citep{b22}. This correlation, 
commonly called Period-Luminosity (PL) relation, was lately renamed as the 
Leavitt law (LL) in honor to its discoverer \citep{b23}.  

This linear statistical correlation is a cornerstone 
in the measurement of the extragalactic distances using 
stellar standard candles. However, in order to obtain an accurate 
calibration of this relation, it is important to establish the effects 
of non-linearities, metallicity and companions on it, as it is 
mentioned by \citet{MF}; as well as other effects such as Cepheids showing 
the Hertzsprung progression and Cepheids exhibiting changes in period. 
A brief literature review of these issues is presented below.

\textbf{i) The non-linearity effect:} the first works in this direction  
began twenty five years after the discovery by Henrietta Leavitt. 
In a study of the shape of the PL relation of nearby galaxies, 
Kukarkin found that this relation has a break around $10$ d (\citealt{Ku}; 
\citealt{b25}). Three decades later, Sandage \& Tammann (1968, 1969) 
found evidence of curvature in the PL relation, and afterward,   
\citet{tsr01} confirmed the break of the PL relation in $10$ d. 
\citet{bb19} confirmed again this  result based on the statistical $F$-test. 
In the last years, the following studies have confirmed the 
non-linearity of the LL without proposing an explanation for it: the testimator 
method \citep{r5}, the approaches of linear regression residuals 
and additive models  \citep{r4},  the multi-phase PL relations
\citep{r2} and the multiple least-squares regression \citep{ag2013}.

\textbf{ii) The metallicity effect:} By using non-linear convective pulsating 
models, \citet{CMM} and \citet{MMF} studied synthetic multiband PL relations  
of populations of Cepheids of different chemical compositions, 
uniformly distributed over the instability strip. 
The masses of these stars  follow the law $dn/dm=m^{-3}$, 
and are distributed in the range  of $5-11$ $M_\odot$.
For a wide range of log $P$, those authors found clear evidence that  
the optical LL is better represented by a quadratic 
relationship than by a linear one, showing a dependence with the metallicity and
the intrinsic width of the instability strip. For IR bands they showed that the 
LL is better fitted by a linear function showing a slightly 
dependence on the metallicity.

\citet{u1} and \citet{b19} found strong evidence 
in favor of the universality of the optical PL relation, 
in the metallicity range from $-1.0$ dex to 
$-0.3$ dex. \citet{g1} and \citet{f1} found that the slopes of the  PL
relations in \textit{VI$W_I$} do not change
significantly between the environments of the Milky Way and the LMC.

\textbf{iii) The companion effects:} 
Physically bound companions to Cepheids
are  difficult to detect at distances of tens of kpc.
To produce a detectable effect on the light curve, 
the  eclipses should be deep enough and 
a significant number of points associated with 
the eclipses should be observed. 
The first condition is reached with an adequate combination of  
radius, luminosity and effective temperature of the primary and secondary
stars, jointly with the appropriate inclination and mass ratio of the system.
These eclipsing systems exhibit light curves similar to those of Cepheids, 
showing besides dispersed points below them, which are a signal of the eclipses 
(\citealt{b2}, Fig. 7). When the Cepheid light curve is subtracted, 
it emerges a typical eclipsing signal (Soszy\'nski et al. 2008, Fig. 3). 

\citet{Num} claim to detect a few Cepheids in eclipsing systems in the
Galactic Bulge and the Magellanic Clouds, using  
a random forest supervised algorithm  
over the MACHO catalog of variable stars. 
However, due to these variables do not show a 
clear period of eclipses, it is difficult to affirm
that they are binaries. It is very likely that
these stars are blended with near neighbors.  \citet{sus} reported few Magellanic Clouds Cepheids that  
could be blended, and also others that could be members of binary systems.

Detection of Cepheid systems whose light curves exhibit eclipsing variations is 
too important: if the Cepheid is a member of a physical binary system, it is
possible to determine dynamical masses, radii and distance with 
a very high accuracy. Until now, spectroscopic 
studies of double-lined eclipsing binary systems in the LMC 
have been made by the Optical Gravitational 
Lensing Experiment (OGLE) in the systems: OGLE-LMC-CEP-$0227$\footnote{The ID 
is given according to the OGLE nomenclature} \citep{Piet}, OGLE-LMC-CEP-$1812$ \citep{Piet11}, 
OGLE-LMC-CEP-$1718$ \citep{Gie14}, OGLE-LMC-CEP-$2532$ \citep{Pil} 
and OGLE-LMC$562.05.9009$ (\citealt{sos12}, \citealt{Gie15})
a system detected in the OGLE Gaia South Ecliptic Pole Field.

\textbf{iv) The Hertzsprung progression effect:} Inspecting  light curves 
of $37$ Cepheids, \citet{Htz} found a relation between the position 
of a bump feature in the light curve and the pulsation period. 
Later studies confirmed that this bump feature is present in the 
light and radial curves of fundamental-mode classical Cepheids
whose periods are in the range of $6-16$ d \citep{Bon}. 
This feature is present in the descending branch
for pulsations in the range of $6-9$ d. For longer periods, this 
feature is observed in the ascending branch and disappears  
for periods longer than $20$ d  (\citealt{Bon}, \citealt{Gas}).
The amplitude bump grows when the period increases, reaching its maximum 
value for periods near to the maximum light around $10-11$ d. For longer 
periods, the  amplitude bump  decreases  until it vanishes \citep{KW}.\\
There are two models that explain some of the 
properties of the observed bump feature.
The first one, called the echo mechanism, proposes 
radial-pressure waves generated in the He II
ionization region \citep{Chr}. The traveling inward wave is reflected
in the core  and reaches the surface one period later, 
leading to the formation of the bump
\citep{Bon}. However, this model has two points to solve: 
the contradiction with the acoustic-ray formalism 
(\citealt{Whit}; Aikawa \& Whitney 1984, 1985) 
and  the difficulty of predicting adequately the Christy wave 
velocity near to the stellar surface (\citealt{Kar}, \citealt{Bon}).

The second model, known as the resonance mechanism,  proposes that 
the feature bump arises from a resonance between the
fundamental mode and the second overtone \citep{Sim}.
By using numerical simulations \citet{Gas} studied the nonlinear saturation 
of the acoustic modes excited by the $\kappa-$mechanism. They found  that 
this $2$:$1$ resonance causes that the bump appears in
the ascending branch for  $P_2/P_0<1/2$.

\textbf{v) The changing period effect:}   
The long-term observations detect several thousand oscillation cycles
for short period Cepheids. This allows to determine, with very high accuracy,  
changing periods. For the long period Cepheids, it is no possible 
to reach this accuracy, because  the data have  
in the  worse  cases of  sampling, tens of cycles. 
There are three scenarios explaining different
characteristics of the period change rates.
The first one proposes that the evolutionary
changes on the stellar structure of Cepheids  
crossing the instability strip are responsible 
of the period change rates (\citealt{Tur}; \citealt{Fdy}).
The second one proposes that the presence of magnetic 
fields is a possible explanation for the 
random changing periods exhibited by some Cepheids.
The last one  suggests that the amplitude and phase
variations of Cepheids are caused by 
an analogous to the Blazhko effect, exhibited by 
RR Lyr\ae{} stars \citep{Mol}.  \citet{sus} reported a 
few Cepheids that could be exhibit this Blazhko effect, 
being the number of these Cepheids larger 
in the LMC than in the SMC. \\

The period changes should produce a shift of the phase 
in the light curve, and as a consequence, a higher scatter 
in the light curve, detectable on years of observations. 
A study of $655$ LMC Cepheids found period 
changes in $18\%$ of the fundamental mode, and in $41\%$ 
of the first overtone (\citealt{Pol}, Table 3). 
A visual inspection of the light curves of these OGLE Cepheids 
shows, for the most of them, a strong scatter of the points 
as a consequence of the shift phase. \\
Since there are a few studies about magnetic 
fields on cool radial pulsating stars, it is not 
clear if the nature of these magnetic fields 
are fossil or are produced as a consequence of 
the stellar pulsation \citep{Wa}. 
Based on a solar-like magneto-convective cycle, \citet{Sto}
could explain the observed period change rates of two 
short period Cepheids: Polaris and V473 Lyr.
As the existence of local magnetic fields on the surface 
of late-type stars causes spots \citep{Str}, \citet{NI} proposed that
convective hot spots can be a possible 
explanation to the random changes in the pulsation 
period detected in the single Cepheid 
observed by the Kepler mission, V1154 Cyg. \\
The first detection of magnetic fields in variable 
stars was made on RR Lyr\ae{} \citep{Bab}.
Measurements of magnetic fields for 
Cepheid stars began later on. In particular, for  $\alpha$ Car 
and $\gamma$ Cyg there were reported values of $700$ G 
\citep{We}, and from $\sim100$ G to $\sim350$ G 
(\citealt{Sev}; \citealt{Plac}), respectively. 
The  spectropolarimetric works over the bright  
bump Cepheid $\eta$ Aql ($V=3.90$, $P=7.17$ d) reported 
controversial results. While \citet{Pla} and \citet{But} 
measured periodic variations of the longitudinal magnetic field
with an amplitude of tens of Gauss, \citet{Wa} found a 
non-significant detection of the longitudinal magnetic field 
at a level of $10$ G.\\

In order to establish whether or not the effects previously described
are influential on the LL, we make a study using statistical 
techniques. We start-off, in the second section,  with a brief description
of the LMC LL and the OGLE-II and OGLE-IV Cepheid data.
The third section is dedicated to review the statistical theory 
of the linear regression analysis, relevant for this work.
The fourth section presents the results of applying statistical models 
in order to obtain optical LL of the LMC and SMC. Finally, our main 
conclusions are given in the fifth section.  

\section{LMC Leavitt law with OGLE data}
OGLE-II and OGLE-IV observations of Cepheid variables of the LMC and SMC 
galaxies were collected with the $1.3$-m Warsaw telescope, 
at Las Campanas Observatory, Chile (\citealt{b2}, \citealt{b3}, \citealt{u15}).
While OGLE-II fundamental mode Cepheid catalogs contain 771 and 1319 stars for 
the LMC and SMC, respectively, OGLE-IV has  a nearly complete collection
(2429 and 2739 for the LMC and SMC, respectively), covering practically 
the whole Magellanic System area with the
time baseline, a little more than five years.
\citep{sus}.

The slope and zero point of the LMC LL in optical \textit{VI}-bands were
computed by \citet{b41} using ordinary least-squares (OLS) regression,
on OGLE-II fundamental-mode Cepheids. Points deviating by more than $2.5 \sigma$
(outliers) were removed by applying the sigma-clipping algorithm \citep{b1}.
Udalski realized that standard deviation of residuals of the LMC data  
was almost two times smaller than that of the SMC. This greater dispersion is 
caused mainly by the spatial distribution of  Cepheids in the SMC bar,  whose 
thickness is placed
along the line-of-sight  with a typical depth of $\sim 0.25$ mag \citep{zz1}.
These facts, and the manner in which the LL of the LMC 
is much better populated for periods longer than $2.5$ d, suggested  
Udalski to adopt as the universal slope value the one obtained for the LMC,  
notwithstanding that the number of Cepheids in the SMC was around twice that of 
LMC. For OGLE-IV LMC Cepheids the slope and zero point of the LL 
were determined in analogous approach, but the SMC slope  was obtained 
independently, using only Cepheids of this galaxy  \citep{sus}.\\

With the aim to  establish a possible statistical 
cause of the non-linearity of the LL, we use the dereddened sample of OGLE-II  .
fundamental mode Cepheids belonging to the LMC and SMC 
galaxies reported by \citet{b2} and \citet{b3},  as well as the nearly complete collection
of OGLE-IV Magellanic Clouds Cepheids reported by \citet{sus}.

\section{Linear regression analysis}
In this section we summarize the topics 
of the statistical theory of linear 
regression analysis, most relevant to this paper.

We start explaining what the outliers and influential points are.
Then, we present the theory of ordinary least square method 
and the conditions to apply it. Next, we describe 
the statistics tools to identify influential points.
Then, we present the specification  test, which verifies if
all variables involved in a statistical problem 
are adequately included and represented by a
proposed model. Following, we present the structural break test,
which looks for unstable parameters.
Finally, we present the robust $M$-regression
that allows to do a linear regression of data with 
no-Gaussian error distribution. The $MM$-regression is applied in order to get 
a 
high breakdown point and fit the
models in presence of influential points that are outliers at the same time.

An extended statistical  discussion of these topics  
can be found in \citet{Mont}. Some applications in the astronomy 
context can be found in \citet{Fei}.

\subsection{OLS statistical theory}
The most common method to estimate 
parameters in a linear regression model is OLS, based on the 
Gauss-Markov theorem. The model is (in matrix notation):
\begin{equation}
\mathbf{y}=\mathbf{X}\boldsymbol{\beta} + \boldsymbol{\varepsilon},
\label{eq1}
\end{equation}
where $\mathbf y$ is the dependent variable vector, 
$\mathbf X$ is the independent variables matrix,  
$\boldsymbol{\beta}$ is the parameters vector 
and ${\boldsymbol{\varepsilon}}$ is the errors vector.

This theorem states that the OLS estimators are the best linear 
unbiased estimators if the expected value of the errors vector is
zero ($E({\boldsymbol{\varepsilon}})=\mathbf{0}$), the variance of the 
errors is constant (${V({\boldsymbol{\varepsilon}})=\sigma^2{\boldsymbol{I}}}$)
and the errors are uncorrelated. 
It can be shown \citep{Mont} that OLS estimators ($\boldsymbol{\hat{\beta}}$) 
are obtained from:
\begin{equation}
\boldsymbol{\hat{\beta}}=(\mathbf{X'X})^{-1}\mathbf{X'y},
\label{eq2}
\end{equation}

Notice that: $\mathbf{0}$ is the nule vector, $\boldsymbol{I}$ the identity 
matrix and $\boldsymbol{A}'$ indicates $\boldsymbol{A}$ matrix is transposed.

Once the model is fitted the adequacy must be checked. That is: 
i) the error term ${\boldsymbol{\varepsilon}}$ has a constant 
variance $\sigma^2{\boldsymbol{I}}$. ii) The errors are uncorrelated. 
These items are the Gauss-Markov theorem assumptions. 
iii) The errors are normally distributed. This is very important 
for making hypothesis testing and estimating confidence intervals. 
iv) All the observations  have approximately the same weight. It
refers to the fact that it is not desirable that  model estimators depend 
more on few observations than on the majority of 
them. This could happen if some observations, called influential points, 
have a disproportional impact in the OLS estimations.
v) There is no specification problems in the model and there is not a 
structural 
break, 
i.e. the parameters are stable. It refers to the correct functional form of the 
model.

After fitting the model, a residual analysis is made to verify its adequacy.
Since the residuals are the difference between 
the original observed values and their fits, they also measure the 
variability of the dependent variable which is not explained by the model.  
Therefore, any violations to 
the assumptions can be detected by analyzing model residuals.  Mainly the 
analysis 
on residuals is looking for evidence of: i) errors come from a distribution 
with heavier tails than normal. Large departures from normality in 
the error distribution mean that the \textit{F-} and \textit{t-} test are not 
longer valid as well as the confidence interval estimations. ii) 
Heteroskedasticity, meaning that the errors do not have constant variance. 
iii) Influential points. iv) Specification problems. v) Autocorrelated errors. 
In this study 
it is not necessary to check if errors 
are correlated because the data are not time dependent, i.e.
they come from a cross-section in the statistical sense.\\
For checking the error distribution a normal probability quantil-quantil 
($Q-Q$) 
plot is used. If all the residuals lie along a line, it means that they come 
from a 
normal distribution. For more details see \citet{Mont}.   
This plot shows if the distribution has 
heavier or lighter tails or if it is 
skewed compared to the normal distribution. 

For checking the assumption that errors have constant 
variance (homoskedasticity) the White's test \citep{White} is used in the 
following equation:
\begin{equation}
\mathbf{y}=\mathbf{X}\boldsymbol{\hat{\beta}}+\mathbf{e},
\label{eq3}
\end{equation}
where $\mathbf e$ is the 
residual 
from the linear regression defined as: 
\begin{equation}
\boldsymbol{\hat{\varepsilon}}=\mathbf{e}=\mathbf{y}-\mathbf{\hat{y}}.
\label{eq4}
\end{equation}
This test does not make the assumption that errors come from a normal 
distribution.  
Its null hypothesis is that $\sigma^2_i=\sigma^2$  for all $i$. 
Since the PL 
relation can be modeled as a simple linear regression, the test uses the 
auxiliary model given by
\begin{equation}
{e^2_i}=\alpha_0+\alpha_1{x_i}+\alpha_2{x_i^2}+{u_i}\quad 
i=1,2,\ldots,n
\label{eq5}
\end{equation}
where $u_i$ is the error term and $n$ is the number of stars.
It can be shown that White's statistic $nR^{2}$ is asymptotic distributed as 
$\chi^{^{2}}$ with g-1 degrees of freedom, where 
$n$ is the number of observations and $R^{2}$ is the regression coefficient of 
determination of the equation (\ref{eq5}). Because the auxiliary model 
(equation \ref{eq5}) has three parameters  ($g=3$) the $\chi^{^{2}}$ has $2$ 
degrees of freedom.\\
\subsection{Outliers and influential points} Outliers are points that have an 
unusual behavior.  
Let us imagine a scatter plot, outliers are data points that unusually 
are located out of the pattern or far from the data cloud. On the other hand,
influential points are those data that have a disproportional impact on OLS 
estimators (slope and zero point).  As a result, the model estimators depend 
more on them than on the majority of data.  
Let us imagine the scatter plot again, a regression line is fitted with 
all points. Then we delete a point and fit the regression 
line once more.  If the regression line obtained after removing the point 
changes 
a lot; the deleted observation is an influential point.  
What is a lot? Statistics helps us in this decision, as we shall describe later.
Now, let us imagine the scatter diagram with a far point from the cloud 
of data points, but located along the regression line.  If we delete it, the 
regression line will not change.  Therefore, the point is an outlier 
but not an influential point, because it affects the $R^2$ coefficient and the 
OLS estimator standard errors, but not the OLS regression estimators.  
On the contrary, if the outlier affects the regression coefficient estimators 
it is also an influential point.
Finally, let us imagine that the point is not an outlier; however, after 
deleting it, the new regression fitted-line is very different from the obtained 
with it.  
This means that it is an influential 
point.  In general, influential points can or cannot be outliers.\\

\textbf{Detection of influential points:} in order to look for influential 
points Cook's distance, \textit{DFFITS, DFBETAS, COVRATIO} statistics are 
 commonly used \citep{Mont}. 
The first of them is defined as 
follows:
\begin{equation}
D_i=\frac{(\boldsymbol{ {\hat{\beta}}}-{\boldsymbol{\hat{\beta}}_{(i)}})' 
\mathbf{(X'X)}  
(\boldsymbol{ 
{\hat{\beta}}}-{\boldsymbol{\hat{\beta}}_{(i)}})}{p\widehat{\sigma}^2},\quad 
i=1,2,\ldots,n
\label{eq6}
\end{equation}
Basically, it measures the square distance between the OLS 
$\boldsymbol{\beta}$ estimate, based on 
all observations and OLS $\boldsymbol{\beta}_{(i)}$ estimate 
obtained after deleting the $ith$ observation, where $p$ in equation 
(\ref{eq6}) 
is the number of parameters in 
equation (\ref{eq3}), i.e. the number of independent variables 
plus the zero point.  Moreover, $\widehat{\sigma}^2$ is an estimate of the mean 
square error, defined as follows:
\begin{equation}
\widehat{\sigma}^2=\frac{\mathbf{{e'e}}}{n-p}.
\label{eq7}
\end{equation}
Observations with large Cook's distance affect OLS estimates of 
$\boldsymbol{\beta}$. To know 
what is a large Cook's distance, $D_i$ is compared to the $F_{0.5(p,n-p)}$ 
distribution \citep{Mont}.  
If  the $ith$ observation has a $D_i > F_{0.5(p,n-p)}$, it is 
considered an 
influential point. 

\textit{DFFITS} measures how large is the influence of the $ith$ observation on 
its 
own estimated $\hat{y}_i$.  In other words, it shows how many standard 
deviations    
$\hat{y}_i$ change due to delete the $ith$ observation.  It is defined as 
follows:

\begin{equation}
DFFITS=\frac{\hat{y}_i-\widehat{y}_{(i)}}{\hat{\sigma}_{(i)}\sqrt{h_{ii}}}, 
\quad i=1,2,\ldots,n
\label{eq8}
\end{equation}

${\hat{y}_{(i)}}$ and ${\hat{\sigma}_{(i)}}$ are the 
fitted values and the 
standard deviation respectively, obtained without the $ith$ observation.  The 
term $h_{ii}$ is the $ith$ diagonal element of the matrix ${\mathbf H}$ that is 
defined as 
follows:
\begin{equation}
\mathbf{H}=\mathbf{X(X'X)^{-1}X'}
\label{eq9}
\end{equation}

The term $h_{ij}$ is the amount of leverage that the $ith$ observation has on 
the $jth$ fitted value. A point is considerate influential when 
$\left|{DFFITS}\right| > 
2  \sqrt{\dfrac{p}{n}} $.
For complementary explanations and references about \textit{DFFITS} and the 
$\mathbf H$ 
matrix see \citet{Mont} and \citet{Dra}. 

Other statistics that detects influential points is \textit{DFBETAS}.
It  indicates how much effect has the $ith$ observation on each $\beta_j$, 
measured in standard deviation units.  In order to apply it, the coefficient 
estimators of 
$\beta_j$ obtained using  all observations are compared with the coefficient 
estimators 
computed excluding the $ith$ observation ($\beta_{j{(i)}}$).  As a result, a 
measure for each 
$\beta_j$ is obtained. The \textit{DFBETAS} statistics is defined as follows:
\begin{equation}
DFBETAS_{j,(i)}=\frac{\hat{\beta}_j-\hat{\beta}_{j(i)}}{\hat{\sigma}_{(i)}\sqrt{
C_{jj}}}\quad 
j=0,1,\ldots,p
\label{eq10}
\end{equation}

where ${\mathbf{C}}=\mathbf{{(X'X)}^{-1}}$, so that $C_{jj}$ is the $jth$ 
diagonal element of the matrix $\mathbf C$ and $\hat{\sigma}_{(i)}$ is the 
square root of the 
regression mean square error fitted without the $ith$ observation. An 
observation is considered 
an influential point if $\left|{DFBETAS}_{j,(i)}\right|>\dfrac{2}{\sqrt{n}}$.\\
The last statistics used to detect influential points is \textit{COVRATIO}, 
that measures how the covariance matrix is affected by the $ith$ 
observation.  It compares the generalized variance   
of the parameters estimators, obtained without the $ith$ observation,  
with that obtained using all observations.\\

The variance-covariance matrix of parameters estimators with all observations 
is:
\begin{equation}
Var(\boldsymbol{\hat{\beta}})=\hat{\sigma}^2(\mathbf{X'X})^{-1}
\label{eq11}
\end{equation}

The generalized variance is the determinant of the equation (\ref{eq11}):
\begin{equation}
GV(\boldsymbol{\hat{\beta}})=\left|{\hat{\sigma}^2(\mathbf{X'X})^{-1}}\right|
\label{eq12}
\end{equation}

Given this, \textit{COVRATIO} is defined as follows:
\begin{equation}
COVRATIO_i=\frac{\left|{\hat{\sigma}^2_{(i)}(\mathbf{{X'}}_{(i)}{\mathbf{X}}_{
(i)}})^{-1}\right|}{
\left|{\hat{\sigma}^2(\mathbf{X'X})^{-1}}\right|}\quad i=1,2,\ldots,n
\label{eq13}
\end{equation}

The $ith$ observation is considered influential if its \textit{COVRATIO} lies 
out of the 
following interval $1-\dfrac{3p}{n}<COVRATIO_i<1+\dfrac{3p}{n} $, 
where $p$ is the number of parameters in the equation (\ref{eq3}) and $n$ is 
the total number of observations. 

The reported LL in this work are obtained by
rejecting from the sample the influential points, and Cepheids exhibiting 
the communalities, explained later. However, two questions 
arise that require detailed answers: 
Why is it important to identify these 
influential points and  give them an statistical treatment? 
Why is it necessary to apply robust techniques to make a 
linear regression instead of applying a well-known algorithm 
as sigma-clipping? In the next paragraphs, we answer 
these questions, illustrating the associated statistical problem.

Outliers are observations with anomalous behavior.
When they are caused by human errors or instrumental failures,
they can be recognized and excluded from the analysis.  Moreover, they can be 
rejected from the 
analysis only if there are strong non-statistical reasons that support such 
decision.
In this work it is possible to reject them based on astronomical reasons.
However, the usual practice of deleting outliers without
making a further analysis looking for reasons that could explain their behavior,
could imply serious consequences in the estimators precision,
because they are adjusted artificially \citep{Mont}.

As a result, when there are not  astronomical  reasons that support excluding
those points and/or there are influential points, robust estimation methods are 
needed.

\subsection{Specification tests} 
To validate that there is no evidence of specification errors in the model, we 
use tests to look if a non-linear combination of the fitted values are 
significant to explain the response variable. 
One of them is the  Ramsey's test given by the equations (\ref{eq15}) to 
(\ref{eq17})
\citep{ram74}. The other one, is a variant proposed by Godfrey and Orme 
that uses only the model given by equation (\ref{eq15}) \citep{God}. 
These auxiliary models and their respectively 
hypotheses used in this work are the following:
\begin{equation}
\mathbf{y}=\mathbf{X}\boldsymbol{\hat{\beta}}+\alpha_1\mathbf{\hat{y}^2}
+\boldsymbol{\varepsilon} 
\label{eq15}
\end{equation}

The null and alternative hypothesis are $H_0:\alpha_1=0$ and 
$H_1:\alpha_1\neq{0}$

\begin{equation}
\mathbf{y}=\mathbf{X}\boldsymbol{\hat{\beta}}+\alpha_1\mathbf{\hat{y}^2}
+\alpha_2\mathbf{\hat{y}^3}+\boldsymbol{\varepsilon}
\label{eq16}
\end{equation}

The null and alternative hypothesis are $H_0:\alpha_1=\alpha_2=0 $ and
$H_1:\mbox{ At least one }\alpha_i\neq{0};\quad i=1,2$

\begin{equation}
\mathbf{y}=\mathbf{X}\boldsymbol{\hat{\beta}}+\alpha_1\mathbf{\hat{y}^2}
+\alpha_2\mathbf{\hat{y}^3}+\alpha_3\mathbf{\hat{y}^4}+\boldsymbol{\varepsilon}
\label{eq17}
\end{equation}

The null and alternative hypothesis are $H_0:\alpha_1=\alpha_2=\alpha_3=0$
and $H_1$: At least one $\alpha_i\neq{0}$; $\quad i=1,2,3$.\\
If any of these tests are significant because there is evidence to reject the 
null hypothesis $H_0$, then there is enough evidence of misspecification.  In 
fact, it means that a non-linear combination of the fitted values are 
significant to explain the variability in the dependent variable, i.e  
the functional form is  incorrect. \\

\subsection{Structural breaks} 
To verify if there are structural breaks, dummy variables are used.
Before explaining what a dummy variable is, it is necessary to give a 
brief explanation about qualitative variables.   Qualitative or categorical 
variables are those that do not have a natural scale of measurement.   
In this study two categorical variables are used.  The first variable indicates 
which data set the observation belongs to when data are divided in two sets, to 
make structural tests to the models.  This variable has two categories: data 
set 
$1$ and data set $2$.  The second variable is the galaxy name where the Cepheid 
is 
located, which also has two categories: LMC and SMC. \\
Dummy variables are used to indicate which category 
from a qualitative variable the observations belong to.  Each dummy variable 
takes two possible values:  $1$ if the observation belongs to the category that 
the dummy variable is representing, and $0$ if not.  If a qualitative variable 
has 
$K$ categories, it can be represented by $K$ dummy variables, one for each 
category. 
 
However, only $K-1$ dummy variables are used in the regression model.  Besides, 
it does not matter which $K-1$ dummy variables are used since 
$\mathbf{H}$ matrix is the same, independently of 
which dummy variable is taken out to fit the model.  In fact, dummy variables 
representing the categories of a qualitative variable are linked between each 
other because when one of them takes a value of $1$ for an observation, the 
others take $0$  for the same observation.  Therefore, when all the $K-1$ dummy 
variables in the 
model take a value of $0$, they are representing the dummy variable that is no 
present in the model. For more details see \citet{Mont}.\\
The process of verifying the existence of structural breaks is made as follows: 
the data are divided in two data sets by dummy variables. Then, a model that 
includes the dummy variables is estimated to verify if the estimated parameters 
associated with dummies are significant or not.  If they are significant, the 
parameters are unstable, thus there are structural breaks.\\
To contrast hypothesis the $P-$value is used.  It means the 
lowest level of significance that leads to reject null hypothesis when it is 
true.  Therefore, if  $P-$value is greater 
than the significance level, there is not enough evidence to reject the null 
hypothesis.  On the contrary, if the $P-$value is less than or equal to the 
significance level there is enough evidence to reject the null hypothesis and 
accept the alternative one, which means the test is significant.

\subsection{\textit{M}-regression} 
Sometimes OLS assumptions are not accomplished; for example, errors have a 
distribution with heavier tails than normal distribution and/or there are 
influential points.   An alternative to fit a linear model in these kinds
of scenarios is a robust regression, in which the residuals  can be defined as 
follows:
\begin{equation}
e_i=y_i-\mathbf{x}'_i\boldsymbol{\beta}\quad 
i=1,2,\ldots,n
\label{eq18}
\end{equation}
It is convenient to scale the residuals by using median absolute deviation:
\begin{equation}
S=  \frac{median \left| {e_i- median(e_i)} \right|}{0.6745}
\label{eq19}
\end{equation}
Therefore the scaled residuals $(u_i)$ can be expressed as follows:
\begin{equation}
u_i=\frac{y_i-\mathbf{x}'_i\boldsymbol{\beta}}{S}\quad 
i=1,2,\ldots,n
\label{eq20}
\end{equation}
If scaled residuals are not dependent between them, and all of them have the 
same distribution  $f(u)$, the maximum likelihood $\boldsymbol{\beta}$  
estimators are 
those that maximize the likelihood function:
\begin{equation}
L(\boldsymbol{\beta})=\prod_{i=1}^n 
f\left(\frac{y_i-\mathbf{x}'_i\boldsymbol{\beta}}{S}\right)
\label{eq21}
\end{equation}
In \citet{Dra} is shown that this condition is equivalent to 
minimize the equation (\ref{eq22}).  Therefore, a class of robust estimator 
that 
minimizes this equation is called $M$-estimator and the regression based on 
it 
is 
called $M$-regression. $M$ comes from maximum-likelihood since function 
$\rho(u_i)$  is 
related to equation (\ref{eq21}) for a proper choice of the error distribution. 
 
Besides, 
$\rho(u_i)$ weights the scaled residuals $u_i$ in the sum:        
\begin{equation}
\displaystyle\sum_{i=1}^n 
\rho\left\{\frac{y_i-\mathbf{x}'_i\boldsymbol{\beta}}{S}\right\}
=\displaystyle\sum_{i=1}^n \rho(u_i)
\label{eq22}
\end{equation}
There are more than one suggestion for $\rho(u_i)$ in the literature, the 
function used in this work is the Tukey's bi-square, with parameter equal to $6$
\citep{KK}. For further information about this function and other choices see 
\citet{Dra}.
To minimize the equation (\ref{eq22}) the first partial derivative of 
$\rho(u_i)$ 
must be calculated:
\begin{equation}
\frac{\partial\rho}{\partial\beta_j}=\psi\left\{\frac{y_i-\mathbf{x}
'_i\boldsymbol{\beta}}{S}\right\}=\psi(u_i)\quad 
i=1,2,\ldots,n,\quad 
j=0,1,2,\ldots,p
\label{eq23}
\end{equation}
As a result, the following equation is obtained:
\begin{equation}
\displaystyle\sum_{i=1}^n x_{i
j}\psi\left\{\frac{y_i-\mathbf{x}'_i\boldsymbol{\beta}}{S}\right\}=0,\quad 
j=0,1,2,\ldots,p
\label{eq24}
\end{equation}
$p$ is the number of parameters in the model.  
Multiplying the equation (\ref{eq24}) by $u_i/u_i$, it is obtained:
\begin{equation}
\displaystyle\sum_{i=1}^n 
x_{ij}w_i \left( \frac{y_i-\mathbf{x}'_i\boldsymbol{\beta}}{S}\right)=0,\quad 
j=0,1,2,\ldots,p
\label{eq25}
\end{equation}
where $w_i= {\psi(u_i)}/{u_i}$
$\forall$ $y_i\neq{\mathbf{{x}}'_i{\boldsymbol{\beta}}}$,
otherwise is $1$. \\
The determination of $M$-estimator is an iterative process, 
thus, the $M$-estimator for $\boldsymbol{\beta}$ is:
\begin{equation}
\boldsymbol{\hat{\beta}_{q+1}}=(\mathbf{X'W}_q\mathbf{X})^{-1}\mathbf{X'W}
_q\mathbf{y}
\label{eq27}
\end{equation}
This iterative process stops when a convergence criteria is reached, 
i.e. when the estimators change less than a preselected amount in the $q+1$ 
iteration.
Notice that the equation (\ref{eq27}) is the weighted least square estimator
where $\mathbf{W}_q$ is a diagonal $n \times n$ matrix of weights whose 
elements 
are $w_i$.  See \citet{Dra} and \citet{Mont} for a wider discussion.

\subsection{\textit{MM}-regression} 
In the context of  $M$-regression it is important 
to check that there are no bad influential points, i.e. 
observations that are outliers in the \textbf{X} space  and influential points 
at the same time.  In this scenario the  $MM$-estimator \citep{Yo} should be 
used because 
it has a high breakdown point (BDP). In a finite-sample the BDP 
is the smallest fraction of  contaminated data causing a divergence of the  
estimator  from the value that it would take if  data  were not contaminated 
\citep{Mont}.\\
The $MM$-estimator combines the asymptotic efficiency of $M$-estimator with 
the high  BDP estimators such as $S-$estimators and Least Treammed Squares.  
$MM$-regression is obtained in three steps. First, a high BDP parameter 
estimators model 
is calculated to compute the residuals of the model. Second, based 
on residual computed in the previous step, a $M$-estimate of scale with high 
BDP is calculated.  Finally, the model parameters 
are estimated using $M$-estimators and the scale estimation computed in the 
previous step. \\
Neither normal errors nor homoskedasticity are assumptions in $M$-regression 
and $MM$-regression.  However, these regressions assume that errors are 
uncorrelated, which is not an issue when data come from a cross-section, 
meaning 
that data are not 
time dependent. \\
Robust regressions are based on asymptotic results. This means they could  
lead to wrong results when they are used with small data sets.  
\citet{Mont} say that small to moderate sample size could be 
less than $50$ points. In our case, the smallest data set has about $400$ observations. 
Hence, we have enough data to apply these techniques without problems.
In particular, OGLE-IV has around two and 
three times of observations for the SMC and the LMC than OGLE-II, respectively.

\section{LMC and SMC Leavitt law regressions}
The statistical theory explained previously are applied for each one of the 
LMC and SMC galaxies. The following models are fitted by OLS: 
\begin{equation}
\mathbf{V_0}=\beta_1+\beta_2\mathbf{LP}+\boldsymbol{\varepsilon}
\label{eq28}
\end{equation}
\begin{equation}
\mathbf{I_0}=\beta_1+\beta_2\mathbf{LP}+\boldsymbol{\varepsilon}
\label{eq29}
\end{equation}
\begin{equation}
\mathbf{W_I}=\beta_1+\beta_2\mathbf{LP}+\boldsymbol{\varepsilon}
\label{eq30}
\end{equation}
$\mathit{LP}$ is $\mathit{\log_{10}P}$ and $P$ is the 
pulsation period of a Cepheid.
In general, these models have violations to the OLS assumptions,
as it will be shown below.

Influential points are data that truly affect the  
linear regression results, although they can or cannot be outliers. 
It is very important that the studies of the LL make 
an appropriate treatment of those points. Deleting a subset of 
influential points does not prevent the appearance of new influential 
points, as will be shown below.\\ 
The sigma-clipping algorithm is an iterative procedure commonly used 
in Astronomy in order to reject the outliers in OLS regression. However, to 
apply the OLS regression its assumptions must be accomplished.
In order to investigate if the OGLE-II data do not fulfill these conditions, 
we make an experiment in V-band using the whole sample of 
$765$ OGLE-II LMC fundamental mode Cepheids. 
We apply the sigma-clipping algorithm  using five iterations 
and an optimum threshold of $2.5\sigma$. These values are the same used in the 
studies of the LL made by \citet{b2} and \citet{ag2013}. 
Simultaneously, we apply the White's test to check if 
the assumption of homoskedasticity is fulfilled.\\
Before the first sigma-clipping iteration, the White's test finds
enough evidence to reject the null hypothesis of homoskedasticity with a 
\textit{P-}value of $0.0165$ at $5\%$ of significance. A number of  $79$ 
influential points are detected using some of the Cook's distance, 
\textit{DFFITS, DFBETAS,} and \textit{COVRATIO}. On the other hand,
by applying the criteria of the sigma-clipping algorithm, $27$ points are 
rejected in the first iteration, not all of them identified as influential 
points.
During second iteration, the White's test finds again enough evidence 
to reject the null hypothesis of homoskedasticity with a \textit{P-}value of 
$0.0483$ at $5\%$ of significance and $76$ influential points are identified.
By using sigma-clipping algorithm $22$ points are rejected. 
In the following three iterations influential points 
are detected again. Taking into account these facts, 
we notice that the main problems of applying the sigma-clipping 
algorithm in the context of the LL are:

i) OLS estimators obtained by  using this procedure   
are not adequate since their precision are adjusted artificially \citep{Mont}.

ii) In the two initial iterations, the models have no a constant error 
variance, i.e., they show heteroskedasticity. In the remaining iterations, 
the models show homoskedasticity, as a result of deleting points using 
the sigma-clipping criterium.

iii) The residuals in each iteration have distributions 
with  heavier tails than the normal distribution.

iv) Influential points are detected on each iteration.

{\textit{Summarizing, rejecting outliers  using the sigma-clipping algorithm is 
not adequate in this case, since there are violations of OLS assumptions. 
Without strong reasons to support the rejection of points, this method is 
invalid in the LL linear regression context.} }

\begin{figure}
\epsscale{.80}
\plotone{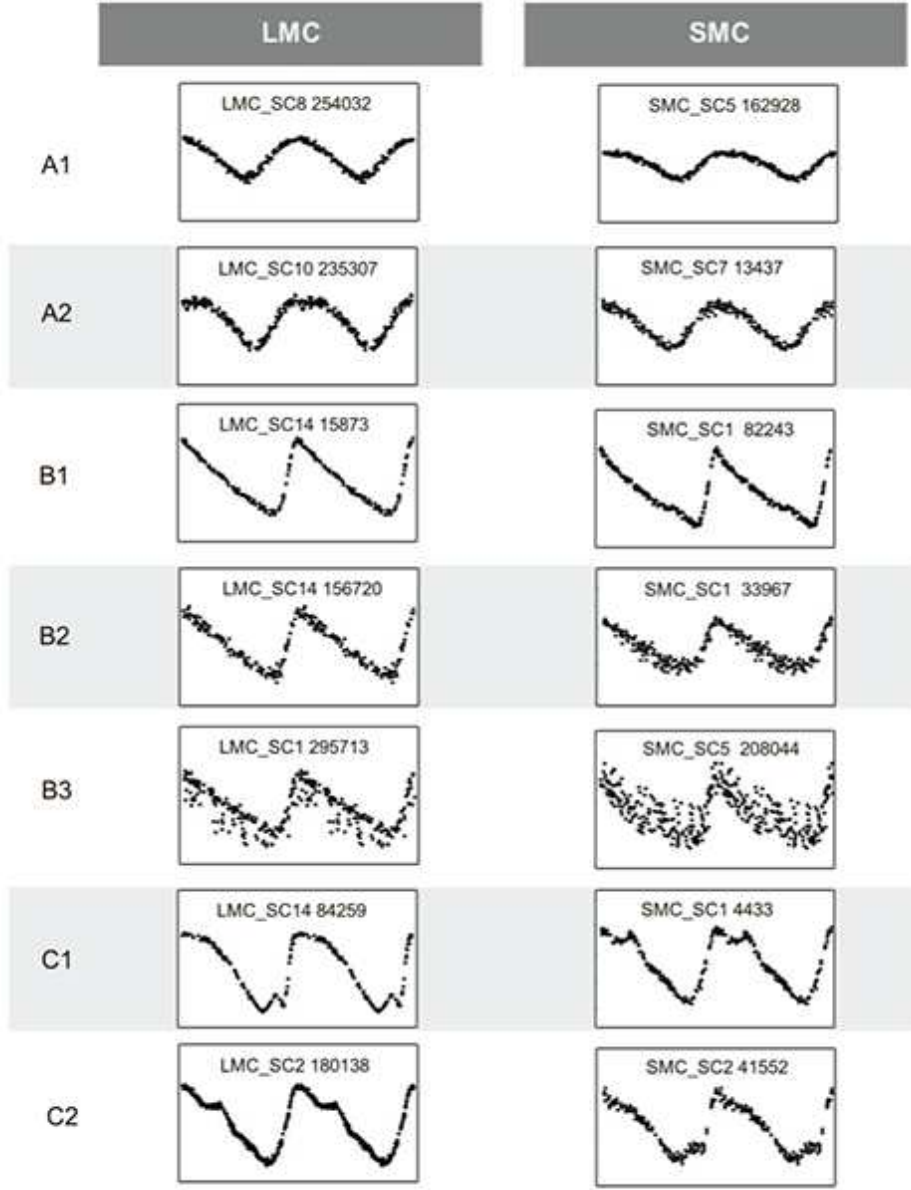}
\caption{Morphological classification of $I$-band light curves 
of selected Cepheids of the Magellanic Clouds. 
The IDs  are given according to the OGLE nomenclature.}
\label{Fig1} 
\end{figure}

As we mentioned before, models (\ref{eq28}) to (\ref{eq30}) have violations to 
the OLS assumptions: for Cepheids belonging to the LMC galaxy, White's test  
(explained in \S3.1) finds enough evidence of heteroskedasticity with 
$P-$values of $0.0165$ and $0.0579$ 
for models (\ref{eq28}) and (\ref{eq29}) at $6\%$ of significance.  
On the contrary, for model (\ref{eq30}), White's test does not find enough 
evidence to reject its null hypothesis because of its $P-$value of $0.4806$, 
that is greater than the same fixed level of significance.
For Cepheids belonging to the SMC galaxy, White's test does not find enough 
evidence to reject the null hypothesis of homoskedasticity with $P-$values of 
$0.3430$, $0.3247$ and 
$0.1599$ for  models (\ref{eq28}) to (\ref{eq30}) respectively, at $5\%$ of 
significance.  \\
Another problem is the influential points that are found in all models in both 
galaxies. As we have mentioned  above, this fact makes necessary to search for 
communalities 
in the influential points to understand the causes of this behavior.
To do that, the $I$-band light curves of all Cepheids in the LMC and SMC  
are visually inspected many times. As a result, we identify two main 
morphological characteristics in them: the dispersion and the shape.
Using these communalities we perform a bi-dimensional classification 
of the Cepheids in some of the following groups: light curves with small (1), 
moderate (2), or large (3) dispersion. And light curves whose morphological 
shapes are sinusoidal-like  (A), sawtooth-like (B) or exhibit 
bumps, i.e. showing the Hertzsprung progression (C). 
Dispersion of light curves of Cepheids classified as 2 and 3 
seem to be caused in several cases by blending, since a visual inspection
of the finding charts of these Cepheids, shows neighbors near to the stars.
Representative examples of this classification 
are shown in Figure \ref{Fig1}. 

The bi-dimensional classification of all OGLE-II LMC and SMC 
fundamental mode Cepheids is given in Table \ref{Tab1}.  
The last column reports our bi-dimensional classification: morphological 
shape is given by A, B or C, and the grade of dispersion 
is measured by the discrete scale: $1$, $2$ or $3$, the last one
being the corresponding to light curves with larger dispersion. An 
appended i letter indicates that this star is detected as 
influential point by at least one of the DFFITS, DFBETAS 
or COVRATIO statistics, computed from $VIW_I$ data. In Table \ref{Tab0},  
the  number of Cepheids belonging to each sub-class
is reported.

\begin{table}
\caption{\textbf{Morphological classification of OGLE-II Cepheids}}
\label{Tab0}
 \centering
\begin{tabular}{@{}ccccccccc}
\\\hline
        & A1 & A2 & B1 &  B2 & B3 & C1 & C2 & Total\\
\hline
LMC & 36 & 9 & 457 & 127 & 7 & 121 & 8 & 765\\

SMC & 10 & 13 & 458 & 650 & 42 & 128 & 5 & 1306 \\ 

\hline
\end{tabular}
\end{table}

\begin{table}[htbp]
  \centering
\caption{\textbf{Bi-dimensional classification of OGLE-II Magellanic Clouds 
Cepheids.}} 
\label{Tab1}
\resizebox{16.5cm}{!}{
    \begin{tabular}{lllll||lllll}\\\hline
\multicolumn{1}{c}{RA} & \multicolumn{1}{c}{DEC } & 
\multicolumn{1}{c}{ID} & \multicolumn{1}{c}{Period} & \multicolumn{1}{c}{Class} 
& \multicolumn{1}{c}{RA} & 
\multicolumn{1}{c}{DEC} & \multicolumn{1}{c}{ID} & \multicolumn{1}{c}{Period} & 
\multicolumn{1}{c}{Class} \\\hline
    05 33 02.35 & -70 15 33.3 & OGLE-LMC\_SC1.25359 & 3.39729 & B1    & 05 27 
57.84 & -69 53 48.5 & OGLE-LMC\_SC3.44391 & 2.53947 & B1 \\
    05 27 43.07 & -70 00 49.7 & OGLE-LMC\_SC3.26910 & 15.84364 & C1i   & 05 27 
34.14& -69 51 22.4 & OGLE-LMC\_SC4.417847 & 7.49402 & C1i \\
    05 24 57.16 & -69 13 32.8 & OGLE-LMC\_SC5.455916 & 9.56961 & C1    & 05 19 
38.13& -69 37 44.6 & OGLE-LMC\_SC7.388032 & 5.66859 & B1 \\
    05 15 37.55 & -69 30 29.4 & OGLE-LMC\_SC8.112083 & 2.98282 & B2i   & 05 12 
06.64& -69 13 06.5 & OGLE-LMC\_SC10.245266 & 3.96572 & B1 \\
    05 42 52.95 & -70 40 11.9 & OGLE-LMC\_SC19.18756 & 3.48887 & B1    & 05 22 
29.04& -70 09 10.2 & OGLE-LMC\_SC21.187856 & 3.16725 & B1i \\
    00 36 55.83 & -73 56 27.2 & OGLE-SMC\_SC1.1     & 14.3816 & C1i   & 00 42 
34.89 & -73 25 37.2 & OGLE-SMC\_SC3.12578 & 1.29508 & B2 \\
    00 44 51.77 & -73 33 56.5 & OGLE-SMC\_SC3.178185 & 3.04334 & B2    & 00 45 
42.73& -72 52 39.7 & OGLE-SMC\_SC4.38887 & 1.91467 & B1 \\  
    00 54 32.93 & -72 30 26.5 & OGLE-SMC\_SC6.324270 & 1.97277 & B2    & 00 55 
18.59& -72 43 12.2 & OGLE-SMC\_SC7.120090 & 1.59513 & B2 \\
    01 01 49.84 & -72 29 55.5 & OGLE-SMC\_SC9.73218 & 3.34991 & B1    & 01 05 
03.87 & -72 51 52.3 & OGLE-SMC\_SC10.78690 & 1.47799 & B2 \\
    01 08 17.81 & -72 44 17.8 & OGLE-SMC\_SC11.75453 & 1.56745 & B1   & 01 09 
04.87 & -72 20 14.5 & OGLE-SMC\_SC11.117272 & 9.15889 & C1i \\\hline 
\tablecomments{The first two columns are the equatorial coordinates of 
the Cepheid ($J2000.0$ equinox).
The third column indicates the OGLE-II ID. The four column 
gives the period. These data were reported for the LMC by  \citet{b2}, and for 
the SMC by \citet{b3}. The last column gives the bi-dimensional class
assigned in this study. The letter i, identifies the influential Cepheids. 
A portion of the table is shown here for guidance regarding its form and 
content. The complete table is available on electronic form at the CDS.} 
\end{tabular}}
\end{table}

When the influential points are correlated 
with our bi-dimensional classification, 
an interesting fact arises:  around one third of the LMC Cepheids 
and one fifth  of the SMC Cepheids, classified 
with a grade of dispersion moderate and large are influential points.  
Taking this fact into account, and in order to avoid violations of OLS 
assumptions, we propose the 
following hypothesis:  If Cepheids with light 
curves of moderate and large dispersion, 
exhibiting characteristics suggesting blending, bumps,
eclipses and period changes, are excluded from the OLS linear
regression, and the remaining data do not present violations 
of OLS assumptions, it is possible to affirm that a cause 
of non-linearities of the LL is due to the inclusion of a fraction 
of these stars, that despite to be Cepheids are  influential points.\\
For testing this hypothesis, we fit the models given by equations 
(\ref{eq28}) to (\ref{eq30}) by OLS using only Cepheids with 
small dispersion. The results confirm the hypothesis, as it will 
be described below. For this reason, we consider appropriate to 
summarize here the communalities of these influential Cepheids:

i) The scatter exhibited by light curves of some Cepheids
classified as A2 or A3 can be associated with 
phase shifts, suggesting changes in the pulsational period.  
This effect produces a dispersion of the mean magnitudes, which
moves the locus Cepheid in the PL relation.

ii) The B2 light curves exhibit a behavior that could be explained
by period changes or blending effects.
The dispersion of these light curves do not allow to directly establish
which of these effects could be present in these types of Cepheids.
In this work it is not clear which of them is more likely to be the best
explanation. However, this point is out of the scope of this paper.
This interesting fact could be investigated in more detail in future works.

A small fraction of the light curves classified 
as B3 have characteristics suggesting eclipses or blending. In the first case, 
the main effect of the companion on the PL relation
is to move the Cepheid locus. Due to that, the total mean brightness 
out-of-eclipse includes the companion contribution, 
adding scatter in some cases up to $1.0$ mag., as it was reported
by \citet{Alc} and \citet{Piet}. The second case is more probable to occur, 
since a visual inspection
of the finding charts of the B3 Cepheids, shows that
most of them are blended. 

iii) Light curves classified as C1 or C2 are bump Cepheids,
i.e. they exhibit  the Hertzsprung progression. The amplitude 
bump affects the Cepheid mean magnitude. This effect is 
stronger for periods near to the maximum light,
shifting the Cepheid position in the PL relation.

In order to use the OLS regression, the White's test is applied.
This test does not find enough evidence to reject its 
null hypothesis of homoskedasticity, with a $5\%$ of significance, since the 
$P-$values associated with each model test are $0.084$, $0.280$ and $0.908$ for 
the LMC, and $0.068$, $0.154$ and $0.440$ for the SMC, respectively.
However, two issues are noted: the first one is again the presence of 
influential points. The second fact is that the residual distributions for 
$VIW_I$  LMC data
(and only for the $W_I$ SMC data) have heavier tails
with respect to the normal distribution. \\
Since there is neither an astronomical nor statistical reason to delete those 
influential points or a fraction of them, a new approach to estimate the 
parameters given in the models (\ref{eq28}) to (\ref{eq30}) is needed instead 
of 
OLS.  
Besides, if these influential points were deleted, new ones may appear. \\
We fit the models by using the $M$-regression. The $\rho(u)$ 
function used is the bi-square. Since no evidence of bad influential points is 
found, we check each model beginning with the slope test, which is a 
robust version equivalent to the $\textit{F}$-test in OLS.
The null and alternative hypotheses are:
$H_0:\beta_2=0$ and $H_1:\beta_2\neq{0}$. \\
For each model fitted in both galaxies, the robust  slope test has a $P$-value 
less than $0.0001$, so at $5\%$ of significance they are significant.  
The LMC $V$-band fitted model (\ref{eq28}) has a $R^{2}$ value
(the coefficient of determination) of $0.6248$, meaning that the model  
explains about the $62\%$ of the variability  of $V$.  For the LMC $I$-band 
model (\ref{eq29}) and the LMC $W_I$ model (\ref{eq30}), they  
explain about $75\%$ and $87\%$ respectively of the observed variability.

We also use the robust Ramsey's test \citep{ram}, given by the following 
equations:
\begin{equation}
\mathbf{y}=\beta_1+\beta_2\mathbf{LP}+\alpha_1\mathbf{\hat{y}^2}+\boldsymbol{
\varepsilon}
\label{eq31}
\end{equation}
The null ($H_0$) and alternative ($H_1$) hypotheses are  
$H_0:\alpha_1=0$ and 
$H_1:\alpha_1\neq{0}$.  
\begin{equation}
\mathbf{y}=\beta_1+\beta_2\mathbf{LP}+\alpha_1\mathbf{\hat{y}^2}+\alpha_2\mathbf
{\hat{y}^3}+\boldsymbol{\varepsilon}
\label{eq32}
\end{equation}
The null and alternative hypotheses are $H_0:\alpha_1=\alpha_2=0$ and
$H_1:$ at least one $\alpha_i\neq{0};\quad i=1,2$. 
\begin{equation}
\mathbf{y}=\beta_1+\beta_2\mathbf{LP}+\alpha_1\mathbf{\hat{y}^2}+\alpha_2\mathbf
{\hat{y}^3}+\alpha_3\mathbf{\hat{y}^4}+\boldsymbol{\varepsilon}
\label{eq33}
\end{equation}
The null and alternative hypotheses are $H_0:\alpha_1=\alpha_2=\alpha_3=0$ and
$H_1:$ at least one $\alpha_i\neq{0};\quad i=1,2,3$. 

In equations (\ref{eq31}) to (\ref{eq33}),  $\mathbf{y}$ is any of the 
magnitude vectors $V$, $I$ or the $W_I$ index.  Moreover, values of $\alpha_i$ 
and $\beta_i$ 
are different for each photometric band. 

This test shows that there is 
not enough evidence to reject $H_0$ due to its $P-$values at $5\%$ of 
significance.  This implies that
there is no evidence of specification problems in the LMC models. 
On the contrary, for the SMC models there is enough evidence to reject
$H_0$ due to $P-$values obtained by the test at the same level of significance.
In particular, when equation (\ref{eq31}) is used, the   LMC (SMC) $P-$values 
are  
$0.544$ ($0.011$) for $V$-band, $0.424$ ($0.006$) 
for $I$-band  and $0.434$ ($0.013$) for the $W_I$ index. 
By using equation (\ref{eq32}), the LMC (SMC) $P-$values are  
$0.832$ ($0.015$) for $V$-band, $0.718$ ($0.006$) 
for $I$-band  and $0.714$ ($0.002$) for the $W_I$ index.
Finally, when equation (\ref{eq33}) is used, the LMC (SMC) $P-$values are 
$0.924$ ($0.013$) for $V$-band, $0.882$ ($0.005$) 
for $I$-band  and $0.709$ ($0.004$) for the $W_I$ index.

\subsection{Structural breaks}
In order to verify if there are structural breaks (explained in \S3.4) 
in model equations (\ref{eq28}) to (\ref{eq30}), we fit the following model:
\begin{equation}
\mathbf{y}=\beta_1+\beta_2\boldsymbol{\delta}+\beta_3\mathbf{LP}
+\beta_4\boldsymbol{\delta}\mathbf{LP}+\boldsymbol{\varepsilon}
\label{eq34}
\end{equation}
Each data set is split in two parts using 
a dummy variable represented by $\delta$, where
\begin{equation}
\delta=\left\{\begin{matrix}
1&\mbox{if P }\leq{3.60 \: (2.57)}\mbox{ d}
\\0&\mbox{if P }>3.60 \: (2.57)\mbox{ d}\end{matrix}\right.
\end{equation}

The LMC (SMC) data are split around $P = 3.60$ $(2.57)$ d, in order to secure 
that each data 
set has enough observations to avoid problems due to asymptotic results in 
robust 
regressions. As a result, each data set has $240$ ($219$) and $247$ ($235$) 
points.

The hypotheses to contrast are: $H_0:\beta_2 = \beta_4 = {0}$ and 
$H_1:$ some $\beta_j\neq{0},\quad j=2,4$. The $\beta_2$ and $\beta_4$ 
parameters are associated with $\delta$ in order to detect structural breaks in 
the zero point $(\beta_2)$ 
and/or the slope $(\beta_4)$ of the LL, if these parameters
are not significant there is not enough evidence of structural breaks.  On the 
contrary, if both or some 
of them are significant there is enough evidence of structural breaks.\\
The obtained LMC $P-$values are $0.886$ for $V-$band, $0.775$ for $I-$band , 
and 
$0.572$ for the $W_I$ index. 
Whereby, no evidence of structural breaks is found, hence the 
parameters obtained using  $M$-estimators are stable at $5\%$ of significance.  

We use the $M$-regression to estimate the LMC LL throughout the
following equations:
\begin{equation}
V_0=17.005-2.683LP+e
\label{eq40}
\end{equation}
\begin{equation}
I_0=16.549-2.930LP+e
\label{eq41}
\end{equation}
\begin{equation}
W_I=15.839-3.315LP+e
\label{eq42}
\end{equation}
Models explain about $62\%$, $75\%$  and $87\%$ of the variability in $V$-band,
$I$-band and $W_I$, respectively. 
The confidence interval at $95\%$ for each $M$-estimator for the LMC zero point  
$(\beta_1)$ and slope $(\beta_2)$ are reported in Table \ref{Tab2}. It is worth 
to say that the estimators for zero point and slope obtained by OLS are 
included 
between the  
confidence intervals obtained by $M$-regression. \\
{\textit{To summarize, the robust tests have shown that there is no evidence to 
have doubts  about 
LMC models adequacy; as a result they are acceptable.}} 
\begin{table}
\caption{\textbf{OGLE-II LMC Leavitt law.}}
\label{Tab2}
 \centering
\begin{tabular}{@{}ccccc}
\\\hline
\hline
        &$\hat{\beta_1}$             &$\hat{\beta_2}$            & ZP           
 
       &  $\eta$ \\
\hline
$V_0$   &17.005  $\pm$  0.064  & -2.683 $\pm$   0.108 &  17.066 $\pm$ 0.021  & 
-2.775  $\pm$  0.031 \\
$I_0$   &16.549  $\pm$  0.044  & -2.930 $\pm$   0.074 &  16.593 $\pm$ 0.014  & 
-2.977  $\pm$  0.021 \\
$W_I$   &15.839  $\pm$  0.029  & -3.315 $\pm$   0.049 &  15.868 $\pm$ 0.008  & 
-3.300  $\pm$  0.011 \\
\hline
\tablecomments{The reported $\beta$ estimators are those given by equations 
(\ref{eq40}) to (\ref{eq42}), and were obtained by using the $M$-regression. 
Confidence radius are reported  at $95\%$. 
The last columns report the zero point and slope
values obtained by \citet{b41} using OLS.}   
\end{tabular}
\end{table}

\begin{table}
\caption{\textbf{OGLE-II SMC Leavitt law.}}
\label{Tab3}
 \centering
\begin{tabular}{@{}ccccc}
\\\hline
\hline
        & $\hat{\beta_1}$             & $\hat{\beta_2}$             
&$\hat{\beta_3}$            & ZP                   \\
\hline
$V_0$   &17.444  $\pm$  0.097  &0.135  $\pm$  0.069  & -2.551 $\pm$   0.147 &  
17.635 $\pm$ 0.031   \\
$I_0$   &17.019  $\pm$  0.074  &0.107  $\pm$  0.053  & -2.869 $\pm$   0.110 &  
17.149 $\pm$ 0.025   \\
$W_I$   &16.353  $\pm$  0.052  &0.063  $\pm$  0.037  & -3.342 $\pm$   0.079 &  
16.381 $\pm$ 0.016   \\
\hline
\tablecomments{ The reported $\beta$ estimators are those given by equations 
(\ref{eq54A}) to (\ref{eq56A}), and were obtained by using the $M$-regression. 
Confidence radius are reported  at $95\%$. The last column report the zero 
point 
value obtained by \citet{b41} using OLS.}
\end{tabular}
\end{table}

\begin{table}
\caption{\textbf{Universal Leavitt law estimators.}}
\label{Tab4}
 \centering
\begin{tabular}{@{}cccc}
\\\hline
\hline
        & $\hat{\beta_1}$               & $\hat{\beta_2}$            
&$\hat{\beta_3}$     \\
\hline
$V_0$   &17.037    $\pm$  0.047  &0.554  $\pm$  0.030  & -2.732 $\pm$ 0.074   \\
$I_0$   &16.590    $\pm$  0.034  &0.540  $\pm$  0.022  & -2.995 $\pm$ 0.054   \\
\hline
\tablecomments{The reported $\beta$ estimators are those given by equations 
(\ref{eq76}) and (\ref{eq77a}), were obtained by using the $M$-regression. 
Confidence radius are reported  at $95\%$.} 
\end{tabular}
\end{table}
For the SMC model the $P$-value obtained by the test of structural breaks 
is less than $0.001$ for all bands, implying enough evidence to reject 
the null hypothesis and accept $H_1$ at $5\%$ of significance; i.e. there are 
structural breaks.

Because of the problems already detected in model equations (\ref{eq28}) to 
(\ref{eq30}) for SMC,
further models analysis is needed using equation (\ref{eq34}). Therefore the 
individual significance of the 
parameters is analyzed.  To do that, the following hypotheses are 
contrasted: 
$H_0:\beta_j = {0}$  and $H_1:\beta_j\neq{0},\quad j=2,4$.
The individual robust test for $\beta_2$ shows it is significant at $5\%$ of 
significance, due to $P-$values of $0.035$ in $V-$band, $0.015$ in $I-$band and 
$0.003$ in 
the $W_I$ index. 
On the contrary, the same test for $\beta_4$ 
is not significant at $5\%$  of significance since associated $P-$values are 
$0.566$ 
in $V-$band, $0.381$ in $I-$band and $0.090$ in the $W_I$ index.
There is not enough evidence to reject $H_0$, i.e. $\beta_4=0$. 
It means that the model has different zero point for each data set but the same 
slope.
Taking into account the specification problems and structural breaks detected 
in models (\ref{eq28}) to (\ref{eq30}), we fit the following model for SMC 
based 
on the previous individual significance 
analysis using (\ref{eq34}):
\begin{equation}
\mathbf{y}=\beta_1+\beta_2\boldsymbol{\delta}+\beta_3\mathbf{LP}+\boldsymbol{
\varepsilon}
\label{eq50A}
\end{equation}
Using OGLE-III Cepheids, \citet{sub} determined the inclination of the SMC 
galaxy's disk in $64^{\circ}$. The 
geometrical distribution of the Cepheids 
across this disk could  explain the behavior of the LL in the SMC, 
observed in equation (\ref{eq50A}): linear regressions with the 
same slope and different zero points for each data set.\\
Model (\ref{eq50A}) is fitted by $M$-regression and no evidence of bad 
influential points is found.  Robust Ramsey's test  is applied 
again to check if there are specification problems in model (\ref{eq50A}). 
The equations and hypotheses of the test are:
\begin{equation}
\mathbf{y}=\beta_1+\beta_2\boldsymbol{\delta}+\beta_3\mathbf{LP}+\alpha_1\mathbf
{\hat{y}^2}+\boldsymbol{\varepsilon}
\label{eq51A}
\end{equation}
The null and alternative hypothesis are:
$H_0:\alpha_1=0$ and 
$H_1:\alpha_1\neq{0}$. 
\begin{equation}
\mathbf{y}=\beta_1+\beta_2\boldsymbol{\delta}+\beta_3\mathbf{LP}+\alpha_1\mathbf
{\hat{y}^2}+\alpha_2\mathbf{\hat{y}^3}+\boldsymbol{\varepsilon}
\label{eq52A}
\end{equation}
The null and alternative hypothesis are:
$H_0:\alpha_1=\alpha_2=0$ and
$H_1:\mbox{At least one }\alpha_i\neq{0};\quad i=1,2$. 
\begin{equation}
\mathbf{y}=\beta_1+\beta_2\boldsymbol{\delta}+\beta_3\mathbf{LP}+\alpha_1\mathbf
{\hat{y}^2}+\alpha_2\mathbf{\hat{y}^3}+\alpha_3\mathbf{\hat{y}^4}+\boldsymbol{
\varepsilon}
\label{eq53A}
\end{equation}
The null and alternative hypothesis are:
$H_0:\alpha_1=\alpha_2=\alpha_3=0$ and $H_1:\mbox{At least one 
}\alpha_i\neq{0};\quad i=1,2,3$. 

The test does not find enough evidence of specification problems 
because the $P-$values obtained at $1\%$ of significance. In particular, for
model (\ref{eq51A}) the SMC $P-$values are $0.933$ for $V$-band, $0.817$ 
for $I$-band  and $0.669$ for the $W_I$ index. For model (\ref{eq52A}), 
SMC $P-$values are $0.370$ for $V-$band, $0.224$ for $I-$band  and $0.040$ 
for the $W_I$ index. For equation (\ref{eq53A}), SMC $P-$values are $0.503$ 
for $V$-band, $0.346$ for $I$-band  and $0.091$ for the $W_I$ index. 

$M$-regression estimations for the SMC LL are shown in the following equations:
\begin{equation}
V_0=17.444+0.135\delta-2.551LP+e
\label{eq54A}
\end{equation}
\begin{equation}
I_0=17.019+0.107\delta-2.869LP+e
\label{eq55A}
\end{equation}
\begin{equation}
W_I=16.353+0.063\delta-3.342LP+e
\label{eq56A}
\end{equation}
These models explain about $75\%$, $82\%$ and $87\%$ 
of the variability in $V$, $I$ and the $W_I$ index, respectively.  The 
confidence intervals at $95\%$ for each $M$-estimator in the equations 
(\ref{eq54A}) to (\ref{eq56A}) are reported in Table \ref{Tab3}.
Besides, it is interesting that the confidence intervals for $M$-estimators 
include the values obtained by OLS, although there are slightly 
differences. \\
{\textit{To summarize, robust tests show specification problems and 
structural break points in the SMC models (\ref{eq28}) to (\ref{eq30}); 
therefore, 
analysis leads to model (\ref{eq50A}) as a better choice.}}    

\subsection{Universality of the Leavitt law}
The universality hypothesis of the LL implies 
that the slope of the linear regression observed in the same filter in 
different galaxies has the same value, and shows a negligible dependence with 
the 
metallicity (\citealt{ag2013}, and references therein.) 
In order to test the universality of the LL, a linear model to fit 
simultaneously the data from the LMC and SMC is proposed.
This model is given by:
\begin{equation}
\mathbf{y}=\beta_1+\beta_2\boldsymbol{\delta}+\beta_3\mathbf{LP}
+\beta_4\boldsymbol{\delta}\mathbf{LP}+\boldsymbol{\varepsilon}
\label{eq73}
\end{equation}
that is fitted in $V$, $I$ and the $W_I$ index after 
deleting the points previously discussed. 
The dummy variable $\delta$ indicates if the observation is from the LMC or the 
SMC: $\delta=1$ if a Cepheid is from the SMC and $\delta=0$ otherwise. 
A number of $454$ Cepheids from the
SMC and 487 from the LMC are used to fit this model.\\
The  model  (\ref{eq73}) is fitted by OLS.  White's test  finds enough evidence 
to reject the null hypothesis of homoskedasticity, at $5\%$ of significance, 
since the $P$-value is $0.017$ in $V$, less than $0.001$ in $I$ and less than 
$0.0001$ in the  $W_I$ index. 
As happened before, OLS regression is not adequate due to  heteroskedasticity 
problems
and the presence of influential points. Besides, the 
residuals distribution has a heavier tail than the normal distribution. 
Therefore, model (\ref{eq73}) is fitted by  $M$-regression using bi-square 
function.  
No evidence of bad influential points is found; thus, the 
process to check the adequacy of the model continues.   \\
The $P-$values of the slope test are less than $0.0001$ in $V$, $I$
and for the $W_I$ index, so the models are significant at $5\%$ of 
significance. 
The hypotheses 
contrasted are: $H_0:\beta_2=\beta_3=\beta_4=0$ and 
$H_1:\mbox{Some }\beta_j\neq{0},\quad j=2,3,4$.  

To test if both galaxies have the same zero point and slope the following 
hypotheses are contrasted: $H_0:\beta_2=\beta_4=0$ and
$H_1:$ Some $\beta_j\neq{0},\quad j=2,4$.\\
The $P-$values obtained by the robust test are less than $0.0001$ in $V$, $I$
and for the $W_I$ index; therefore, there is enough 
evidence to reject the null hypothesis at $5\%$ of significance. Hence, 
the individual significance of each parameter must be analyzed.  To do that, 
the following hypotheses are contrasted again: $H_0:\beta_j=0$ and
$H_1:\beta_j\neq{0},\quad j=2,4$.\\
The individual significance robust test for $\beta_2$, shows it is significant 
at $5\%$  of significance, since its associated $P-$values  are less than 
$0.0001$  in $VI$ and the $W_I$ index. The same test for $\beta_4$  
is not significant at $5\%$ of significance because its associated $P-$values 
are $0.456$ and $0.145$ in $V$ and $I$, respectively. This means that
the model has different zero points, as we expect due to the galaxies are
at different distances,  but it has the same slope. 
As a result, the final equations fitted by the $M$-regression
in $V$ and $I$ bands, after deleting $\beta_4$ are: 
\begin{equation}
V_0=17.037+0.554\delta-2.732LP+e
\label{eq76}
\end{equation}
\begin{equation}
I_0=16.590+0.540\delta-2.995LP+e
\label{eq77a}
\end{equation}
Confidence radius are reported in Table \ref{Tab4}.
Models (\ref{eq76}) and (\ref{eq77a}) make an adjustment to the zero point 
when Cepheids are from the SMC but have the same slope for both galaxies at 
$5\%$ of significance.  Moreover, they explain about the $78\%$ and $81\%$ 
of the variability in $V$ and $I$, respectively.

The individual significance robust test for  $\beta_4$ in the $W_I$ index 
is significant at $5\%$  of significance because its associated $P$-value is 
$0.0014$. It implies
that each galaxy has its own regression line with different zero point and 
slope.  
It explains about $87\%$ of the $W_I$ variability. Following equation gives the 
model fitted for the $W_I$ index:\\
\begin{equation}
W_I=15.841+0.594\delta-3.318LP-0.129\delta LP+e
\label{eq78a}
\end{equation}
The confidence intervals at $95\%$ for estimated parameters in equation 
(\ref{eq78a}) are:
$15.801\leq\beta_1\leq{15.880}$ for zero point,
$0.548\leq\beta_2\leq{0.639}$ for zero point adjustment when Cepheids are from 
SMC,
$-3.384\leq\beta_3\leq{-3.251}$ for slope, and
$-0.209\leq\beta_4\leq{-0.05}$ for slope adjustment when Cepheids are from 
SMC.\\
{\textit{As a general result of this work, robust tests have shown that the LL 
of the LMC and SMC are universal in $V$ and $I$ bands. There are enough evidence of two 
parallel egression lines, with the same slope at $5\%$ of significance but different 
zero points, as it is shown by models (\ref{eq76}) and (\ref{eq77a}).
However, for the $W_I$ index, it has been shown that there are not a universal
LL: the LMC and SMC have two completely different regression 
lines, as it is shown by equation (\ref{eq78a}).}} 

\subsection{Analysis with OGLE-II and OGLE-IV data}    
Based on the experience obtained by fitting the models 
(\ref{eq28}) and (\ref{eq29}), they are fitted for all $VI$ mean magnitudes of 
fundamental mode Cepheids observed by the OGLE-II project. We also fit these 
models using the OGLE-IV fundamental mode  Cepheids, reported by \citet{sus}.
It is worthy to say that the published OGLE-IV mean magnitudes 
are not corrected for extinction.
All following models are fitted with
the $MM$-regression because of its high BDP of $25\%$.
The Least Treammed Squares and Tukey function are used. \\
As no evidence is found of skewness in the models for both galaxies, we proceed 
with the analysis. Beginning with the LMC, model slope tests are significant, 
as it can be seen in the $P-$values obtained: less than $0.0001$ in $V-$  and $I-$band
(OGLE-II, OGLE-IV).  Then, the test proposed by \citet{God} is performed 
obtaining $P-$values of  $0.302$ (OGLE-II) 
and $0.033$ (OGLE-IV) in $V-$band and $0.186$ (OGLE-II) and $0.0003$ (OGLE-IV) 
in $I-$band; therefore, there is evidence of specification problems in OGLE-IV 
$I-$band model at $1\%$ of significance.  This could be due to blending 
problems and also to the lack of extinction 
correction in OGLE-IV data. To perform stability tests, the data 
are split in two data sets around $P=3.85$ d (OGLE-II)
and $P=3.5877$ d (OGLE-IV), to have about half of observations 
in each set, by dummy variables.  No evidence of structural breaks
is found, because the $P-$values obtained, by testing 
$\beta_2=\beta_4=0$ vs the hypothesis that  at least one of them is not $0$ 
in equation (\ref{eq34}), are $0.601$ 
(OGLE-II) and $0.453$ (OGLE-IV)  in $V-$band, 
and $0.810$ (OGLE-II) and $0.267$ (OGLE-IV) in $I-$band. Models explain about 
$61\%$ and $68\%$ of variability  in $V-$ and $I-$bands respectively (OGLE-II); 
and $54\%$ and $65\%$ of variability in $V-$ and $I-$bands respectively 
(OGLE-IV). The estimated parameters of the models and their confidence intervals
are reported in Table \ref{Tab6}. Results for $I-$band are shown only  
for information purposes because of their specification problems. 
\begin{table}
\caption{\textbf{Optical LMC Leavitt law.}}
\label{Tab6}
 \centering
 \resizebox{16cm}{!}{
\begin{tabular}{@{}ccccccc}
\\\hline
        &\multicolumn{2}{c}{OGLE-II}  &\multicolumn{2}{c}{OGLE-IV} 
&\multicolumn{2}{c}{Soszy\'nski et al.}\\
        & $\hat{\beta_1}$             &$\hat{\beta_2}$            
&$\hat{\beta_1}$                  & $\hat{\beta_2}$ & ZP & Slope \\
\hline
$V$   &17.029  $\pm$  0.041  & -2.732 $\pm$   0.063 &  17.436 $\pm$ 0.025  & 
-2.701  $\pm$  0.038 & 17.438 $\pm$ 0.012 & -2.690 $\pm$  0.018 \\
$I$   &16.553  $\pm$  0.029  & -2.947 $\pm$   0.043 &  16.823 $\pm$ 0.018  & 
-2.922  $\pm$  0.029 & 16.822 $\pm$ 0.009 &  -2.911 $\pm$  0.014 \\
\hline
\tablecomments{The reported $\beta$ estimators are those given by equations 
(\ref{eq28}) to (\ref{eq29}). They were obtained by using the $MM$-regression. 
Confidence radius are reported  at $95\%$. The last columns report the zero 
point and the slope values obtained by \citet{sus} using the sigma-clipping 
algorithm and OLS over OGLE-IV data. The obtained OGLE-II and OGLE-IV
$MM$-estimators in $V-$band and their respective confidence intervals include at 
$1.0\sigma$ the OLS estimators reported by \citet{b41} and \citet{sus}.}   
\end{tabular}
}
\end{table}

The OGLE-IV LMC data were also split in around $P=10$ d by equation 
(\ref{eq34}). No evidence of structural break is found in the slope, 
because the $P$-value obtained is $0.705$. However, the zero point 
adjustment is significative. As a result, there is evidence of 
two parallel regression lines.  \\
The SMC slope tests for models (\ref{eq28})
and (\ref{eq29}) are significant because the $P-$values 
obtained are less than $0.0001$ for $V-$ and $I-$bands (OGLE-II, OGLE-IV).  
However, these models show 
problems.  Godfrey and Orme test finds enough evidence of specification 
problems because of $P­-$values obtained  are less than $0.0001$ for $V-$ and $I-$bands 
(OGLE-II, OGLE-IV). To perform stability tests, data are split around $2.0039$ 
d (OGLE-II) and $1.9429$ d (OGLE-IV)  to have about half of 
observations in each data set. Enough evidence of structural breaks is found:  
$P-$values are less than $0.0001$ in both bands (OGLE-II), 
for testing $\beta_2=\beta_4=0$ vs the hypothesis that at least one of them 
is not $0$ in equation (\ref{eq34}). Besides, when individual 
significance tests are performed for $\beta_2$ and $\beta_4$, $P-$values 
obtained are less than $0.0001$ for $\beta_2$ in both bands; $0.0322$ and 
$0.0082$ for $\beta_4$ in $V-$band and $I-$band respectively.  As a result, 
in both bands there are two regression lines at $5\%$ 
of significance (OGLE-II). \\
Unlike SMC OGLE-II models, there is not enough evidence to support that OGLE-IV 
slope model parameters are unstable with $P-$values $0.784$ and $0.597$  in 
$V-$ and $I-$bands respectively, while zero point adjustment, $\beta_2$ 
(equation (\ref{eq34})), is still significant with $P-$value less than 
$0.0001$ in both bands.  As a result, stability tests show that there 
are two parallel regression lines in both bands. Based on the individual 
significance analysis using model (\ref{eq34}) for $V-$ and $I-$bands, 
new models are fitted. However, Godfrey and Orme test  shows that specification 
problems persist. Nevertheless, only for information purposes, 
we report the slope estimated values at $95\%$ confidence 
intervals for the OGLE-IV data:$-2.780\pm0.055$ for $V-$band and 
$-3.024\pm0.046$ for $I-$band. \\ 
{\textit{Summarizing, the LMC models fitted with OGLE-II and OGLE-IV 
data are adequate except for the $I-$band model (OGLE-IV). The  OGLE-II models 
adequacy could be explained because only $\sim 36\%$ of Cepheids in the  
LMC has intermediate and large dispersion. 
Therefore, as Cepheids with small dispersion are the 
largest fraction, excluding intermediate 
and large dispersion Cepheids has no significant impact 
on the estimated slope value, because confidence interval
for slope ($\hat{\beta_2}$), showed in Table \ref{Tab6}, 
includes the slope estimation ($\hat{\beta_2}$) showed 
in Table \ref{Tab2}. Moreover, LMC fitted models with all OGLE-IV data have 
confidence intervals for slope that contain estimation obtained by using the models fitted  
with all OGLE-II data (see $\hat{\beta_2}$ values in Table \ref{Tab6})}} .\\
{\textit{On the other hand, there is enough evidence showing 
that SMC models are not adequate. Unlike LMC, the largest fraction of OGLE-II 
SMC data $(\sim 64\%)$ are intermediate and large dispersion 
Cepheids; as a result, when models are fitted with all data, 
those Cepheids are majority in the sample.
The presence of a large fraction of intermediate and large dispersion Cepheids
in the SMC galaxy can be explained by the geometry 
of this galaxy, as it was discussed in the section $4.1$. 
It is interesting that SMC models improve when they are fitted with OGLE-IV 
data because, despite of specification problems, there is not  
enough evidence of unstability of the slope parameter in both bands. }}

\section{Conclusions}
Our main conclusions can be summarized as follows: 

By applying the traditional OLS and robust statistical linear regression models to 
optical OGLE-II and OGLE-IV data, it is possible to affirm that 
the problems of break and non-linearity of the LL disappear for the 
LMC when  $M$- or $MM$-regressions are used. In that case, the LL is obtained 
without slope breaks, as it is shown in equations (\ref{eq76}) and (\ref{eq77a}), and it is not needed 
to exclude or reject points from the data sample.
The models in this case are adequate and do not present specification problems 
or structural breaks, except in the OGLE-IV $I$-band, probably due to
the lack of extinction correction  among others causes. 
These facts allow to conclude that the non-linearity of the LMC LL is a 
consequence of using a non-adequate statistical method. \\
For the SMC, despite to use the robust regressions, specification, structural breaks and adequacy 
problems are found. This could be caused in part by blending problems 
due to the geometry of this galaxy, 
that could be generating a large fraction of 
intermediate and large dispersion Cepheids,
as it was discussed in  section $4.1$. This astronomical reason seems to imply 
that Cepheid variables in galaxies with elongated structures
like this of the SMC
are not appropriate to fit a reliable LL. In particular, for the SMC, we found 
that excluding the intermediate and large dispersion Cepheids, i.e. stars 
that exhibit the communalities of influential Cepheids,
leads to avoid the adequacy and specification problems.  
This could be indicating that this kind of Cepheids
really has influence in the fitting of the LL for 
galaxies with similar geometry to this of the SMC. However, 
from a statistical point of view, there is no reason to exclude these stars from the analysis.\\

The OLS method has violations to its assumptions   
for the LMC and SMC galaxies, with OGLE-II and OGLE-IV data,
as it was discussed at the beginning of section $3.1$.
In particular, two important problems are worthy of mentioning here.
The first one is the fact that rejecting outliers 
using the sigma-clipping algorithm is not adequate 
because the precision of the parameters is adjusted 
artificially, besides there are violations of OLS assumptions.
Without strong non-statistical reasons to support the rejection of 
points, this method is invalid in the LL linear 
regression context. The second problem is the presence 
of influential points, that implies the necessity to find
communalities in them. In this work we found that a fraction of 
Cepheids with a grade of moderate and large dispersion 
are influential points. The communalities exhibited by 
them suggest in their light curves bumps, i.e. Cepheids showing the 
Hertzsprung progression, or eclipsing variations, or  
evidence of period changes. The exclusion 
of these influential Cepheids from the sample using 
OLS regression was made to solve adequacy problems;  however, new influential points 
appear that imply again violations to the OLS assumptions. \\

Although the obtained  $M$-estimators (slope and zero point) and their respective confidence 
intervals include the OLS estimators reported by \citet{b41}, 
the relevance of using the $M$-estimators
is that they allow to make a reliable statistical inference, 
since none of their assumptions are violated, unlike the OLS estimations.\\
When Cepheids from the LMC and SMC are combined to fit a single model, 
in each one of the $V-$ and $I-$bands, the linear relations remain to be valid,  
after deleting the Cepheids already discussed.
The $M$-model shows that each galaxy has its own regression line differing only 
in the zero point.  
This implies that the OGLE-II Leavitt law is universal in the range of 
metallicities of the Magellanic Clouds.
On the contrary, the combined model for $W_I$ shows that each galaxy has its 
own linear regression with the expected difference in zero point, but
with unexpected difference in slope. This result has already been 
reported in the literature, but its causes are not yet clear.

LMC models fitted by using  all OGLE-II and 
OGLE-IV data by $MM$-regression are adequate, 
but OGLE-IV $I-$band has specification problems. These problems could be 
due to blending and the lack of extinction correction in these data. There is 
no evidence of unstability  in their zero point and slope parameters. 
On the other hand, despite of the  specification  problems in SMC models 
for OGLE-II and OGLE-IV data, there is not enough evidence of unstability in the slope 
parameters when they are fitted with OGLE-IV data that have more observations 
compared to OGLE-II. However, when OGLE-IV data are split in $P=10$ d, 
there is evidence of two parallel regression lines. \\
It is clear from our results that influential Cepheids do not have an important effect 
on the LL for the LMC. However, they could have impact 
in the case of galaxies with similar geometry to this of
the SMC. The effect of these Cepheids in the measurement of 
distances to these galaxies, will be studied in a forthcoming paper.\\
Finally, based on the  results of this work, we suggest to 
use  the LL relations given by equations
(\ref{eq76}) and (\ref{eq77a}) as a universal law for 
the SMC and LMC galaxies. We also suggest to use the 
slope and zero point for $V$-band, obtained for
LMC using OGLE-IV data,  shown in Table \ref{Tab6}. 
For the SMC using OGLE-II without the influential Cepheids, 
we suggest to use the equations (\ref{eq54A}) and (\ref{eq55A}).\\

\acknowledgments
Authors thank the referee, Dr. Pawel Pietrukowicz, 
for his important recommendations and  suggestions 
that have improved this manuscript.
AGV, BES and JRM acknowledge support from Facultad de Ciencias, Universidad
de los Andes, through Proyecto Semilla. SVD acknowledges the Physics 
Department, Universidad de los Andes, for the  fellowship supporting the postdoctoral
position held within the Astronomy group.

\end{document}